\documentclass[table]{article} %

\usepackage{iclr2024_conference,times}

\usepackage{amsmath,amsfonts,bm}

\def\eqref#1{equation~\ref{#1}}

\def\1{\bm{1}}

\DeclareMathAlphabet{\mathsfit}{\encodingdefault}{\sfdefault}{m}{sl}
\SetMathAlphabet{\mathsfit}{bold}{\encodingdefault}{\sfdefault}{bx}{n}

\definecolor{codegreen}{rgb}{0,0.5,0}
\definecolor{codeblue}{rgb}{0.25,0.5,0.5}
\definecolor{codegray}{rgb}{0.6,0.6,0.6}
\definecolor{OliveGreen}{rgb}{0,0.6,0}

\definecolor{lightblue}{rgb}{0.8,0.85,0.96}
\definecolor{heavyblue}{rgb}{0.46,0.62,0.89}
\definecolor{heavyblue}{rgb}{0.56,0.68,0.91}
\definecolor{lightorange}{rgb}{0.97,0.89,0.81}
\definecolor{heavyorange}{rgb}{0.92,0.70,0.46}
\definecolor{lightgreen}{rgb}{0.86,0.91,0.83}
\definecolor{heavygreen}{rgb}{0.61,0.75,0.51}

\usepackage{hyperref}

\hypersetup{
    colorlinks,
    citecolor=codeblue,
    filecolor=codeblue,
    linkcolor=codeblue,
    urlcolor=codeblue
}

\usepackage{bookmark}
\usepackage{url}
\usepackage{graphicx}
\usepackage{wrapfig}

\usepackage{amsfonts}
\usepackage{amsmath}
\usepackage{amssymb}
\usepackage{booktabs}
\usepackage{multirow}
\usepackage{multicol}
\usepackage{colortbl}
\usepackage{listings}
\usepackage{amsmath}
\usepackage{bbm}
\usepackage{enumitem}
\usepackage[capitalise]{cleveref}
\usepackage{xspace}

\usepackage{booktabs}

\usepackage{algorithm}
\usepackage{algorithmic}

\usepackage[titles]{tocloft}
\usepackage{listings}
\usepackage{authblk}

\makeatletter
\renewcommand\AB@affilsepx{, \protect\Affilfont}
\makeatother
\usepackage{pifont}%
\newcommand{\cmark}{\textcolor{black}{\ding{51}}}
\newcommand{\xmark}{\textcolor{gray}{\ding{55}}}

\newcommand{\method}{\textsc{DEI}\xspace}
\newcommand{\baseline}{\textsc{\textbf{DeiBase}}\xspace}

\definecolor{kellygreen}{rgb}{0.3, 0.73, 0.09}
\definecolor{alizarin}{rgb}{0.82, 0.1, 0.26}
\usepackage{pifont}%

\usepackage{threeparttable} %
\usepackage{caption}
\usepackage{subcaption}
\usepackage{tcolorbox} %

 \title{Diversity Empowers Intelligence: Integrating Expertise of Software Engineering Agents}

\author[1,2]{Kexun Zhang}
\author[1]{Weiran Yao}
\author[1]{Zuxin Liu}
\author[1]{Yihao Feng}
\author[1]{Zhiwei Liu}
\author[1]{Rithesh Murthy}
\author[1]{Tian Lan} 
\author[2]{Lei Li}
\author[1]{Renze Lou}
\author[1]{Jiacheng Xu}
\author[1]{Bo Pang}
\author[1]{Yingbo Zhou}
\author[1]{Shelby Heinecke}
\author[1]{\\Silvio Savarese}
\author[1]{Huan Wang}
\author[1]{Caiming Xiong}

\affil[1]{Salesforce AI Research}
\affil[2]{Carnegie Mellon University}

\iclrfinalcopy %
\begin{document}

\maketitle

\renewcommand{\thefootnote}{\fnsymbol{footnote}}
    \footnotetext[1]{Code, data and leaderboard results at: \href{https://salesforce-research-dei-agents.github.io}{\texttt{salesforce-research-dei-agents.github.io}}}
    \footnotetext[2]{Contact: \href{mailto:kexun@cmu.edu}{\texttt{kexun@cmu.edu}}, \href{mailto:weiran.yao@salesforce.com}{\texttt{weiran.yao@salesforce.com}}}
\renewcommand{\thefootnote}{\arabic{footnote}}
\vspace{-5mm}

\begin{abstract}
Large language model (LLM) agents have shown great potential in solving real-world software engineering (SWE) problems. The most advanced open-source SWE agent can resolve over 27\% of real GitHub issues in SWE-Bench Lite. However, these sophisticated agent frameworks exhibit varying strengths, excelling in certain tasks while underperforming in others. To fully harness the diversity of these agents, we propose \textbf{\method} (Diversity Empowered Intelligence), a framework that leverages their unique expertise. \method functions as a meta-module atop existing SWE agent frameworks, managing agent collectives for enhanced problem-solving. Experimental results show that a \method-guided committee of agents is able to surpass the best individual agent's performance by a large margin.
For instance, a group of open-source SWE agents, with a maximum individual resolve rate of 27.3\% on SWE-Bench Lite, can achieve a 34.3\% resolve rate with \method, making a $25\%$ improvement and beating most closed-source solutions. Our best-performing group excels with a 55\% resolve rate, \textit{securing the highest ranking} on SWE-Bench Lite. Our findings contribute to the growing body of research on collaborative AI systems and their potential to solve complex software engineering challenges.
\end{abstract}

\section{Introduction}
\label{sec:intro}

Recent advancements in large language models (LLMs) have transformed software engineering (SWE) and other domains. Originally developed as chatbots~\citep{schulman2022introducing,openai2024gpt4technicalreport}, LLMs have evolved into the core of AI agents, capable of understanding and generating human-like conversations, as well as autonomously executing actions in both real-world and digital environments. SWE agents, a specialized subset of these AI agents, integrate these capabilities with software engineering tools and techniques for tasks like code generation, automated testing, and project management, aiming to identify and resolve practical software issues~\citep{zhang2024autocoderover}.

\begin{figure}[ht]
    \centering
    \includegraphics[width=\linewidth]{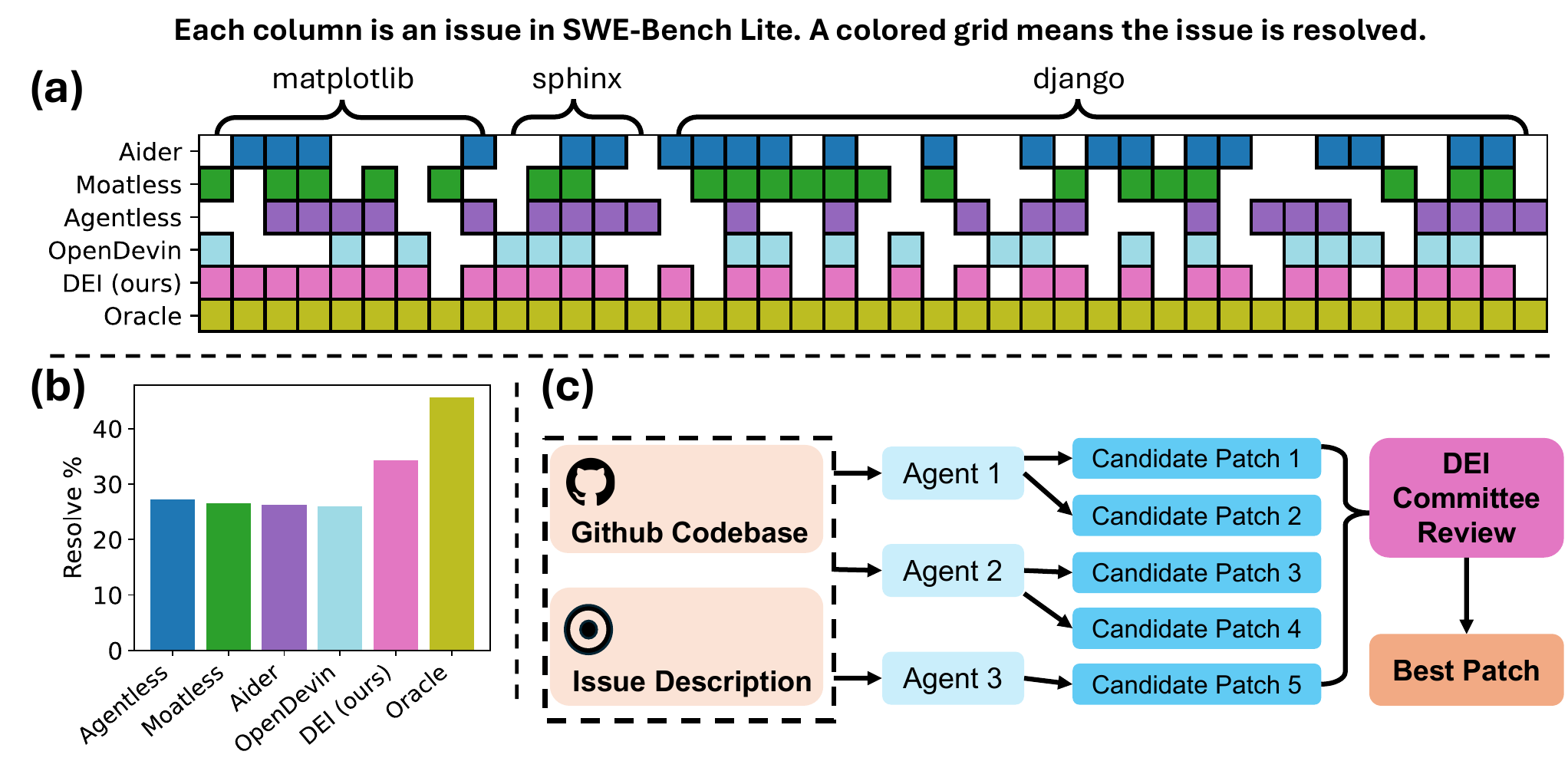}
    \vspace{-14pt}
    \caption{Different SWE agents (Aider, Moatless, Agentless, OpenDevin) resolve very different sets of issues (the colored girds in Fig \textbf{1a}), despite having similar resolve rates (Fig \textbf{1b}).
    Our proposed \method Committee takes candidates patches and tries to select the best, oracle choice (Fig \textbf{1c}), improving the resolve rate significantly to be better than any single agent in the committee.}
    \label{fig:problem}
\end{figure}

In this paper, we study one specific task of SWE agents -- resolving real-world GitHub issues based on their descriptions. Automatically fixing a bug in a code repository is an extremely challenging task that involves navigating extensive codebases, understanding complex function interactions, detecting subtle errors, and generating the correct fix patch. The large action space of SWE agents, together with long trajectories, inevitably result in the diversity of Github issue solutions, as shown in \autoref{fig:problem}. We have observed that different SWE agents resolve very different sets of issues (the colored girds in \autoref{fig:problem}\textbf{a}), despite having similar resolve rates (\autoref{fig:problem}\textbf{b}).
This is probably due to different skill sets of SWE agents.
For instance, OpenDevin~\citep{wang2024opendevin} explicitly instructs the LLM to first replicate the bug in an issue and executes its replication in a development workspace to provide feedback for its generated patches, but other agents like Moatless Tools \citep{orwall2024moatless} and Agentless \citep{orwall2024moatless} do not actually execute code in the issue-specific repository.

\begin{center}
\vspace{-5pt}
	{\bf \textit{A garden's beauty never lies in one flower. Diversity in all its forms is the path to greatness.}}
\end{center}

Similarly, the trend in the SWE agent community reflects this diversity—no single agent framework dominates in all capabilities. It is the flourishing variety within this community that sparks new ideas and leads to the development of better agents. 

The variety in SWE agent capabilities inspires us to develop \method, \underline{\textbf{D}}iversity \underline{\textbf{E}}mpowered \underline{\textbf{I}}ntelligence, a framework that leverages the strengths of diverse agents. \method aims to harness these varied skills to tackle a broader range of problems more effectively with a multi-agent ensemble system and a re-ranking pipeline, as showcased in \autoref{fig:problem}\textbf{c}. \method functions as a meta-module that can be integrated with any existing agent framework, enabling scalable management and collaboration among agents to form a more powerful multi-agent software engineering organization.

We evaluate \method on 7 groups of candidate agents on SWE-Bench Lite.
3 of the 7 are different runs of a single open-source SWE agent.
The other 4 are different agents that are on the SWE-Bench Lite leaderboard, including a group that contains only open-source agents.
Through experiments, we find that different agents show a great level of diversity in the issues they resolve: a group of agents with an average resolve rate of 26.6\% can collectively solve 54.3\% of issues if there is an oracle reviewer that can consistently select the best candidate.
\method, as a first step towards harnessing such diversity, can improve the group's resolve rate to 34.3\% ($\uparrow 25\%$), suggesting that LLMs are great code reviewers.
These findings mirror the benefits of diversity in the tech industry, where diverse perspectives and skills lead to greater innovation and problem-solving capabilities.

To summarize, our contributions are the following:
\begin{itemize}[leftmargin=*] \itemsep0em
    \item  For the first time, we comprehensively evaluate the diversity of solutions provided by SWE agents, revealing significant differences in the types of GitHub issues resolved by various agents, despite similar overall resolve rates. These findings suggest a substantial potential to improve overall performance by more effectively leveraging the diverse expertise of these agents together.
    \item This paper introduces \method, a multi-agent meta-policy module designed to harness the diversity of SWE agents and seamlessly facilitate collaboration among agents with different specialties. By employing a multi-stage rating and re-ranking pipeline, \method consistently improves issue resolution, demonstrating a 25\% performance boost on the SWE-Bench Lite leaderboard.
\end{itemize}

\section{Related Work}
We review the work in fundamental language agent architecture, recent developments for SWE agents, and multi-agent or ensemble methods in this section. 

\paragraph{Fundamental Language Agent}
Pioneering AI agent methods along this line of work include ReAct~\citep{yao2023react}, Reflexion~\citep{shinn2023reflexion}, CodeAct~\citep{wang2024executable}, etc., in which ReAct interprets the user query, generates functional API calls, and gets the tool outputs in real time; Reflexion further appends failed trial experience into the memory, enabling effective retrials to prevent repetitive errors. CodeAct~\citep{wang2024executable}, instead of generating function calls, uses code generation to consolidate AI agents’ actions into a unified action space.

\paragraph{Software Engineering Agent}

We present the SWE agents which have disclosed the technical details on the SWE-bench Lite leaderboard. Alibaba Lingma Agent~\citep{ma2024understand} constructs a repository knowledge graph to represent code and dependencies, using a Monte Carlo tree search-based strategy for repository exploration, and generates patches to address real-world GitHub issues. AutoCodeRover~\citep{zhang2024autocoderover} adds advanced code search tools, such as abstract syntax trees, and spectrum-based fault localization to the agent for enhancing context understanding and issue resolution. Code-R~\citep{chen2024coder} chooses a multi-agent framework with pre-defined task graphs to resolve Github issues. Agentless~\citep{xia2024agentless}, is a simplified two-phase approach for solving software development problems. It focused on localization and repair without relying on LLMs to make decisions, highlighting the potential of straightforward techniques in autonomous software development. OpenDevin~\citep{wang2024opendevin} is a hub of community-contributed agents including CodeAct~\citep{wang2024executable}, browser agent, GPTSwarm~\citep{zhuge2024language}, and task-specific micro agents. Finally, SWE-agent~\citep{yang2024swe} developed agent-computer interface that consists of LM-friendly commands and environment feedback to enable LM agents to autonomously use computers to
solve software engineering tasks.

\paragraph{Multi and Ensemble Agents}
Recent works observe that organizing multiple specialized AI agents~\citep{hong2024metagpt,li2023camel,liu2024agentlite} enable the task decomposition ability of an agent system, which improves the task-resolving performance. 
Current multi-agent frameworks are categorized into three types based on their execution patterns. 
Firstly, static agent working flow~\citep{wu2024stateflow,babyagi}, which pre-defines the agent execution flows and ignites agent transitions via specified conditions. 
Controlling a multi-agent system with pre-determined states is robust, though losing flexibility in terms of unseen states or conditions. 
Secondly, ensemble via group chatting~\citep{wu2023autogen,hong2024metagpt,wang2024mixture,chen2023agentverse}.
This is built upon an environment where
multiple agents send messages to each other in a group channel such that their thoughts are ensembled.
Variants of group chatting includes debating~\citep{liang2023encouraging,chan2023chateval} and model-wise ensembling~\citep{wang2024mixture}.
Last but not least, hierarchical task assignment~\citep{liu2024agentlite,liu2023bolaa}.
Organizing multi-agent in a hierarchical structure benefits the top-down task decomposition and thus enables efficient multi-agent collaboration.

\section{Integrating Expertise of SWE Agents}

\subsection{Background}

\textbf{Resolving issues in SWE-Bench.}
One important task in software engineering is to resolve issues raised by developers.
SWE-Bench curates instances of this task by collecting successfully resolved issues from open-source repositories on Github.
Each instance in SWE-Bench consists of a textual issue description, a version of the repo just before the issue was resolved, and (hidden) unit tests that went from fail to pass after the human-written patch.
To resolve an instance, the model is required to generate a patch that can pass these unit tests.

\textbf{SWE Agents.} In this paper, we use the term ``SWE agents''\footnote{According to our definition, SWE-agent \citep{yang2024swe} is an instance of SWE agents, and Agentless \citep{xia2024agentless}, despite the name, is another.}
to refer to \textit{any} LLM-based system that generates patches to solve issues in a code base, e.g., an instance in SWE-Bench.
While the specific implementation varies, a typical SWE agent usually gives their underlying LLM several tools in the form of callable functions to navigate through the code base, find relevant context, edit files, and run tests.
The workflow of SWE agents often involves multiple LLM calls, each taking some or all outputs from previous steps as input.

\subsection{Diversity of SWE Agents}

We consider two types of diversity: \textit{intra-agent diversity} and \textit{inter-agent diversity}.

\textit{Intra-agent diversity} is defined as the degree to which different runs of the same agent solve different problem instances.
It is most likely from the non-determinism of the underlying LLM due to sampling in decoding and mixture-of-experts architecture \citep{chann2023non}.
Since the workflow of SWE agents involves multiple steps and LLM calls, a slight difference in an earlier step can easily propagate and result in significant differences in the final outcome.

\textit{Inter-agent diversity} is defined as the degree to which different agents solve different problem instances.
Besides sharing the potential causes of intra-agent diversity, inter-agent diversity is also largely because of differences in agent design, including different tools, workflows, and prompts.

\subsection{Approach}

\subsubsection{SWE Agent Problem Formulation}

We formulate the SWE agent problem under the \textit{contextual Markov decision process} (CMDP) framework \citep{hallak2015contextual}, represented by the tuple \( \mathcal{M} = (\mathcal{S}, \mathcal{C}, \mathcal{A}, \mathcal{R}, \mathcal{P}, p_0, \rho) \). Here, \( \mathcal{S} \) denotes the state space, which encompasses all possible states the agent could encounter, such as the current status of files. The context space, \( \mathcal{C} \), includes relevant repository information and issue descriptions. The action space, \( \mathcal{A} \), represents all potential actions or tools the SWE agent can utilize, such as \texttt{search} or \texttt{editing}. The context-dependent reward function, \( \mathcal{R} : \mathcal{S} \times \mathcal{A} \times \mathcal{C} \to \mathbb{R} \), assigns scores based on the actions taken by the agent. For instance, the reward is high if the agent successfully addresses an issue, while it is low if the action results in new bugs in the repository.
The context-dependent transition function, \( \mathcal{P} : \mathcal{S} \times \mathcal{A} \times \mathcal{C} \to \Delta(\mathcal{S}) \), defines how the state of the repository or information changes following a specific action. The context-dependent initial state distribution is denoted by \( p_0 : \mathcal{C} \to \Delta(\mathcal{S}) \), and \( \rho \in \Delta(\mathcal{C}) \) represents the context distribution.

Given the initial context \( c \sim \rho \) and initial state \( s_0 \sim p_0(\cdot \mid c) \), at each time step \( t \), the agent follows a policy \( \pi : \mathcal{S} \times \mathcal{C} \to \Delta(\mathcal{A}) \) to select an action \( a_t \sim \pi(s_t, c) \) and receives a reward \( \mathcal{R}(s_t, a_t, c) \). The environment then transitions to the next state \( s_{t+1} \sim \mathcal{P}(\cdot \mid s_t, a_t, c) \), providing the agent with a new state observation.
As the iteration progresses to time \( T \), a sampled trajectory \( \tau := \{s_t, a_t, r_t\}_{t=0}^{T} \) is obtained. The objective of an SWE agent is to maximize the cumulative reward along the trajectory, which is captured by the value function:
\begin{equation}
    \max_{\pi} V^{\pi}(\rho) = \max_{\pi} \mathbb{E}_{\tau} \left[ \sum_{t=0}^{T} \mathcal{R}(s_t, a_t, c) \mid c \sim \rho; \pi \right]
    \label{eq:objective}
\end{equation}
\subsubsection{Our Framework: Diversity Empowered Intelligence (DEI)}
Many efforts have been made to implement sophisticated agent systems that aim to achieve the objective described in Eq.~\ref{eq:objective}. However, as discussed in Section \ref{sec:intro}, these systems often exhibit varying levels of effectiveness across different contexts. It is challenging to devise a single agent that can consistently perform well across all possible contexts.

Formally, suppose there are \(N\) agent policies, denoted as \(\{\pi_1, \pi_2, \dots, \pi_N\}\), where each policy is tailored to address a specific context \(\{\rho_1, \rho_2, \dots, \rho_N\}\). The union of these contexts is a subset of the entire context space, i.e., \(\rho_1 \cup \rho_2 \cup \dots \cup \rho_N \subseteq \rho\). For each agent policy \(\pi_i\), the objective is:
\begin{equation}
    \pi_i = \max_{\pi} \mathbb{E}_{\tau} \left[ \sum_{t=0}^{T} \mathcal{R}(s_t, a_t, c) \mid c \sim \rho_i; \pi \right].
\end{equation}
However, an agent policy \(\pi_i\) may perform poorly in a different context \(\rho_j\) (where \(j \neq i\)). To address this limitation, we propose our framework: Diversity Empowered Intelligence (DEI). The DEI framework leverages the strengths of each agent in their respective contexts to enhance overall performance across all contexts.

We introduce a meta-policy, denoted as \(\pi_{\text{DEI}}\), which aims to optimally select among the available agent policies based on the context. The goal of \(\pi_{\text{DEI}}\) is defined as:
\begin{equation}
    \pi_{\text{DEI}} = \max_{\pi} \mathbb{E}_{c \sim \rho} \left[ \mathbb{E}_{\tau} \left[ \sum_{t=0}^{T} \mathcal{R}(s_t, a_t, c) \mid c; \pi(c) \right] \right],
\end{equation}
where \(\pi(c)\) denotes the selection of the optimal agent policy from \(\{\pi_1, \pi_2, \dots, \pi_N\}\) based on the observed context \(c\). By dynamically choosing the most suitable agent policy for each context, the DEI framework seeks to maximize the expected cumulative reward across all possible contexts.

\subsubsection{\baseline: A Simple yet Effective Implementation}

We present \baseline, a simple yet powerful implementation of the DEI framework, tailored for SWE-Bench like problems. The context in the setup includes the repository, along with relevant files and issue descriptions. The meta-policy's action space consists of the final patches generated by different agent frameworks, each specialized in addressing various aspects of the problem.

\baseline utilizes a Large Language Model (LLM) as a code review committee. The LLM evaluates candidate patches by analyzing the state of the code base before and after the proposed changes, in conjunction with the contextual information from the issue descriptions. It produces detailed explanations for each patch, justifying the modifications based on the identified issues, the context, and the specific changes made.

While other methods of code review and scoring, such as rule-based approaches, can be incorporated into our framework, the use of an LLM-based committee offers a unique advantage. LLMs often excel at evaluating solutions when evaluation is easier than generation. \baseline thus serves as an effective baseline for LLM-based SWE evaluation, highlighting potential performance variations among diverse SWE agents and showcasing the capabilities of our method.

\begin{figure}[ht]
    \centering
    \includegraphics[width=\linewidth]{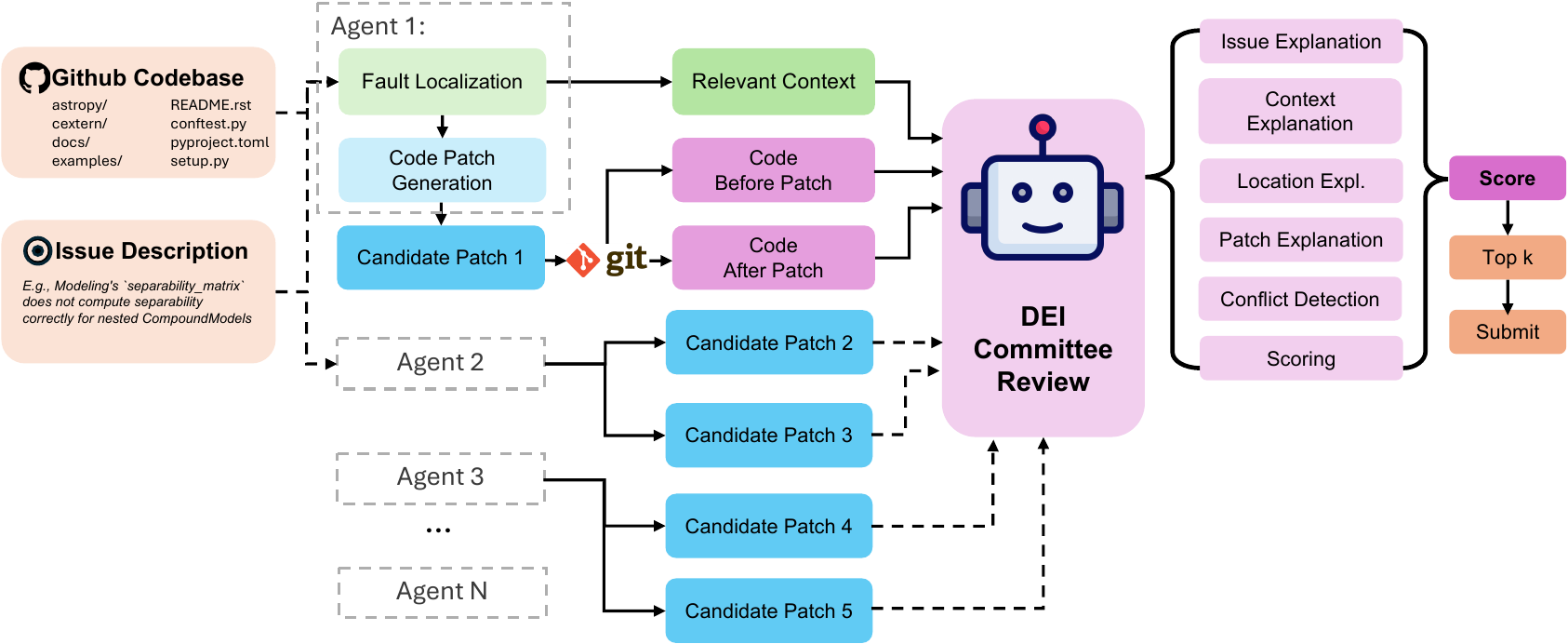}
    \caption{\textbf{Framework Overview.} \method first examines the code base before and after a candidate patch, along with other relevant contexts. Then, it generates an explanation for the issue, the context, and the patch and tries to justify the patch. With its own explanation, it scores the candidate patches and picks the top-scoring ones as more likely to be correct.}
    \label{fig:arch}
\end{figure}

As demonstrated in \autoref{fig:arch}, \baseline is given multiple candidate patches for a single issue.
These patches might be from running a single SWE agent multiple times or running multiple SWE agents.
\baseline gives each candidate patch a score and then selects the top-scoring candidates as the patches most likely to work.

\textbf{Step 1: Input Construction.} Four inputs are given to \baseline for each patch: the issue description itself, relevant context (code snippets identified by an SWE agent as relevant to the issue), code before the patch, and code after the patch.
The form of inputs reflects two design choices. First, the entire repository is often too large to fit directly in the context limit of LLMs, so we use the relevant context instead to save token costs and help the model focus.
Second, the format of a patch is not the easiest for an LLM to read as it switches back and forth between the pre-change code and the changed code, so we give the code before and after the patch separately to the model for easier understanding.
In practice, we directly use the relevant code spans identified by Moatless Tools, an open-source SWE Agent \citep{orwall2024moatless}.
There might be potential ways of improving the quality of relevant code spans by making them specific to both the issue and the candidate patch, rather than solely dependent on the issue itself.

\textbf{Step 2: Explanation Generation.}
To help the model better ``understand'' the patch before scoring, we instruct it to generate various explanations regarding the patch in a specified order.
The order is decided so that the earlier explanations can also help the later ones.
We describe each explanation in the order they are generated here:
1) \textit{Issue explanation} explains what the issue is about and what problem it may be causing.
2) \textit{Context explanation} explains how and why each relevant code span (there might be many of these) is relevant to the issue.
3) \textit{Location explanation} explains if and why the patch is modifying the correct part of the code that's faulty.
4) \textit{Patch explanation} explains if and how the patch is fixing the issue.
5) \textit{Conflict detection} is about checking whether the patch conflicts with other relevant code snippets.
We explicitly instruct the model to refer back to the earlier explanations while generating the later ones.

\textbf{Step 3: Patch Scoring.} Based on its own explanations, the model is asked to give the candidate patch a score of 1 to 10. We give the model detailed rubrics of what violations/mistakes lead to higher score deductions and what should only be considered minor violations.
For example, if the model finds the modification location to be wrong, it is considered a serious mistake.

\section{Experiments}

We aim to answer two research questions with our experiments: 1) How diverse are LLM-based SWE agents in terms of intra- and inter-agent diversity? 2) To what extent can \method harness the diversity and increase the performances of these SWE agents?

\subsection{Experiment Setup}

\subsubsection{Benchmark and Agents}

\textbf{Benchmark}. We conduct our experiments on SWE-Bench Lite, a 300-instance subset sampled from the full SWE-Bench for providing a more self-contained evaluation of functional bug fixes~\citep{jimenez2024swebench}.
Compared to the full SWE-Bench, SWE-Bench Lite has significantly more submissions on the leaderboard for us to conduct a more comprehensive analysis of inter-agent diversity.

\textbf{Agents.}
For intra-agent diversity, we consider three well-performing open-source agents on the SWE-Bench Lite leaderboard: Agentless \citep{xia2024agentless}, Moatless \citep{orwall2024moatless}, and Aider \citep{gauthier2024aider} by running them 10 times with the same parameters.
For inter-agent diversity, we consider 10 agents (open-source or not) that have similar resolve rates, all between 26.0\% and 31.0\% on the leaderboard by directly using their submitted patches to the SWE-Bench issues.
For the evaluation of \baseline on different agents, we consider 3 groups of agents that form different \method Committees, including one group consisting of only open-source agents.
For the evaluation of \baseline on multiple runs of a single agent, we use the generations of the three aforementioned agents -- Agentless, Moatless Tools, and Aider.
More details can be found in Appendix \ref{app:agents}.

\subsubsection{Evaluation Metrics}

We use the same set of metrics for both intra- and inter-agent diversity as these metrics are defined for multiple candidate solutions without requiring them to come from the same candidate:

\textbf{Resolve rate} measures how good a SWE agent is. It is defined as the percentage of issues resolved by the agent. We measure both single SWE agents and \method with it to see how much \method helps.

\textbf{Union@k} measures the \textit{best case} performance had the agents been perfectly consistent by counting the number of problems solved by any of the $k$ solutions.
In the ideal case where the agents are perfectly consistent, Union@k should be the same as Union@1.
Union@k can also be considered as the case where we have an oracle reward function $\mathcal{R}^{\text{oracle}}$ that always selects the correct candidate.

\textbf{Intersect@k} measures the \textit{worst case} performance by computing the number of problems solved by all $k$ solutions.
The assumption is that a problem is only consistently solved if it's always solved.
\citet{yao2024tau} calls this metric pass\^{}k.
Intersect@k can also be considered as the case where we have an adversarial reward function $\mathcal{R}^{\text{adv}}$ that tries to pick an incorrect candidate if there is one.

\textbf{Average@k} measures the \textit{average case} performance by computing the average number of problems solved. It corresponds to the case of a random reward function $\mathcal{R}^{\text{random}}$ that uniformly samples a candidate solution for each problem.

\textbf{n@k} measures the performance of any reranking mechanism by computing the number of problems solved by $n$ chosen submissions from a total of $k$ samples.
The better a reranking mechanism is at telling good solutions from bad ones, the higher n@k is.
Note that for an oracle that always picks the correct solution over incorrect ones, n@k is the same as Union@k.
For a random reranker that picks a random solution uniformly, n@k is the same as Union@n.
In our case, we evaluate $n=1$.

Our research questions can be answered by the gaps between these metrics.
\textbf{Union@k - Intersect@} measures how diverse the agents are, while \textbf{n@k - Average@k} measures how much \method helps in selecting the correct candidate from these agents.
Note that the order -- in which different runs are added -- matters as $k$ gets larger, especially when the $k$ candidate solutions come from $k$ different agents.
In our experiments, we add candidate solutions from the single agent according to the order they are generated, while we add solutions from different agents in a fixed order (see Appendix \ref{app:agents}).

\subsection{Main Results}

\subsubsection{Research Question 1: How diverse are LLM-based SWE agents?}

\begin{figure}[H]
     \centering
     \begin{subfigure}[b]{0.47\textwidth}
         \centering
         \includegraphics[width=\textwidth]{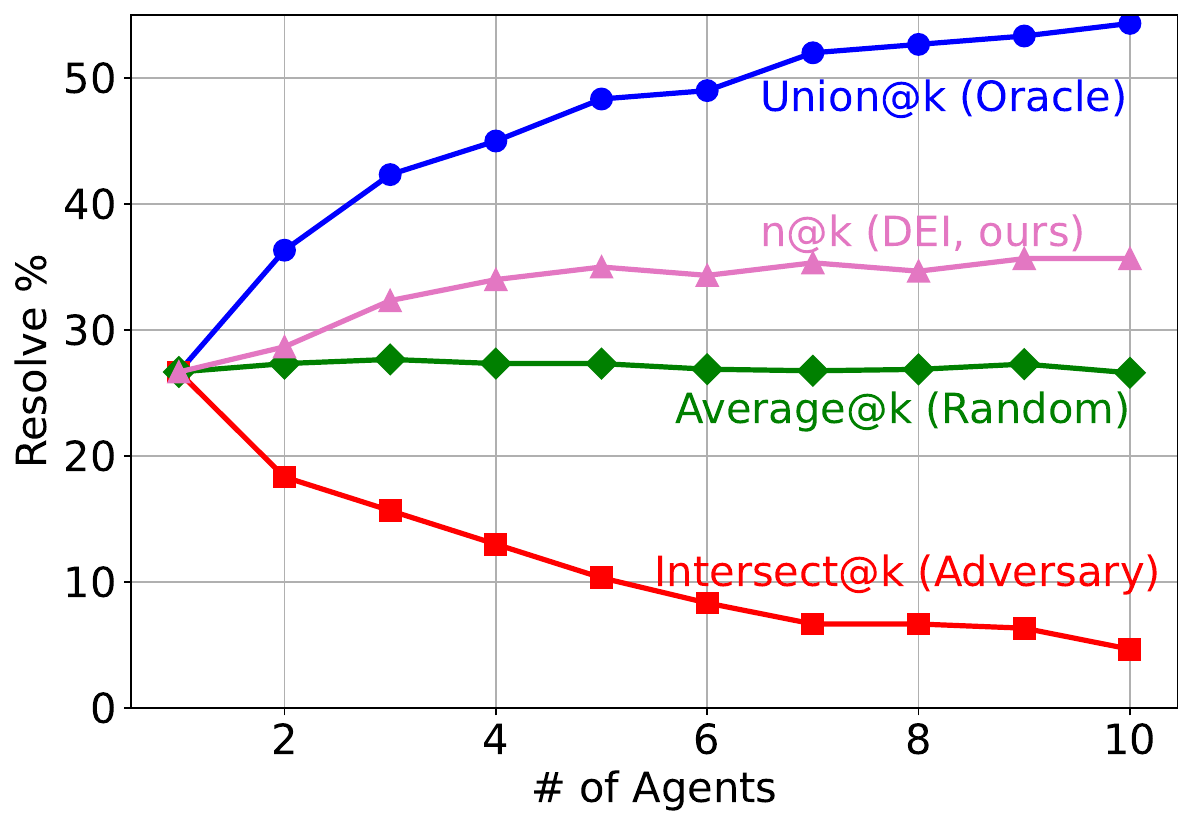}
         \label{fig:multiagent}
     \end{subfigure}
     \begin{subfigure}[b]{0.47\textwidth}
         \centering
         \includegraphics[width=\textwidth]{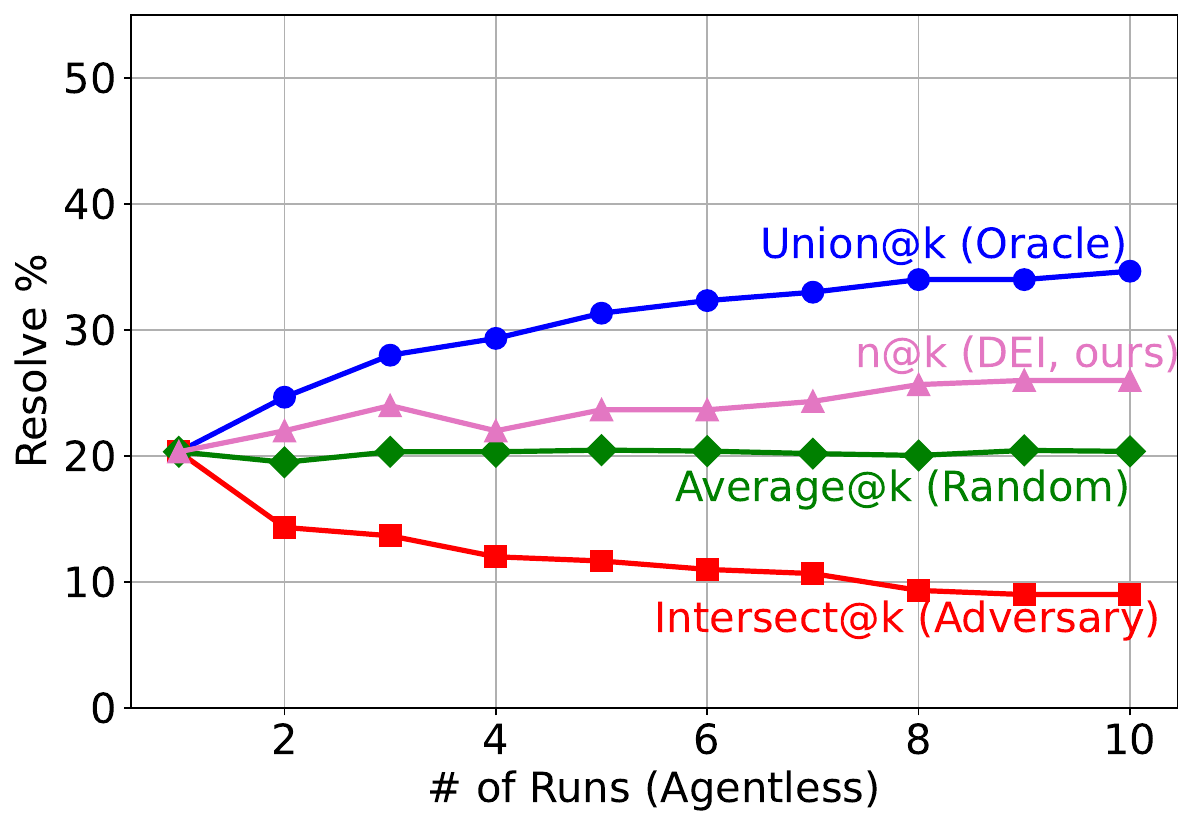}
         \label{fig:gugu}
     \end{subfigure}
    \vspace{-10pt}
    \\
    \begin{subfigure}[b]{0.47\textwidth}
         \centering
         \includegraphics[width=\textwidth]{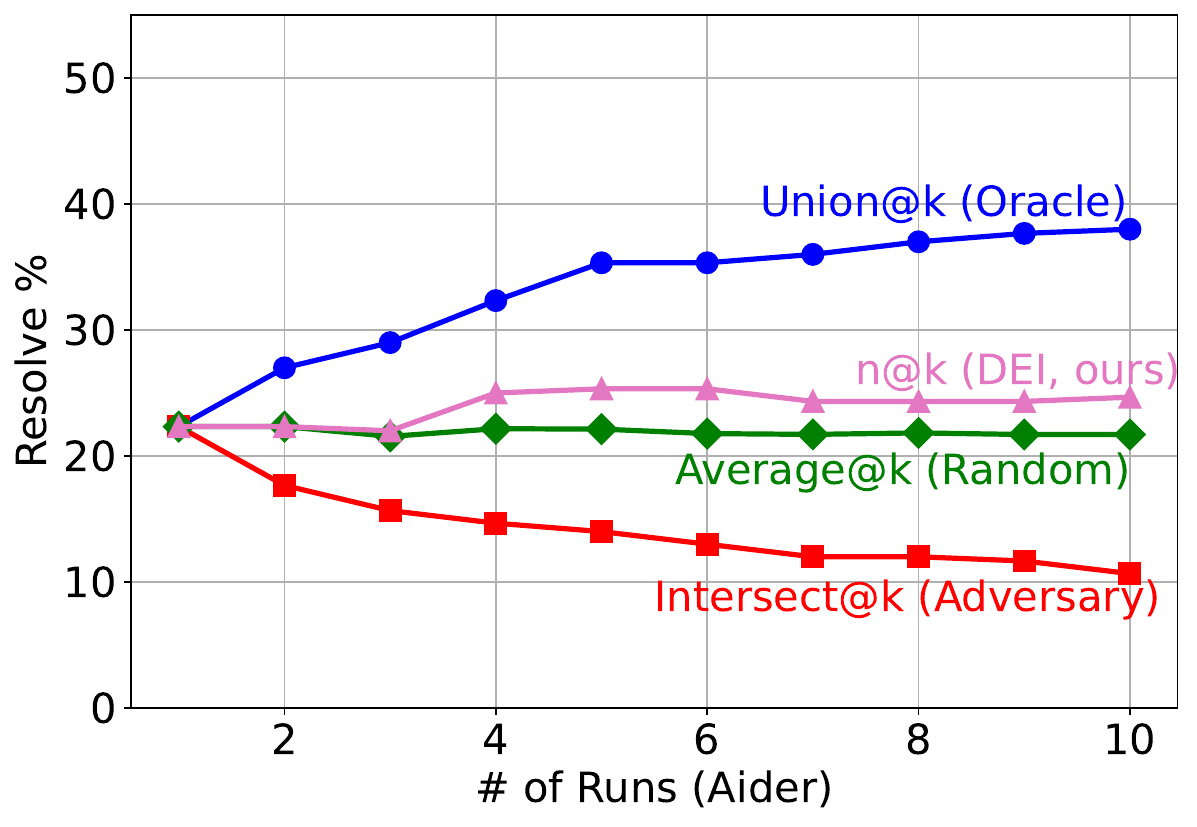}
         \label{fig:three sin x}
     \end{subfigure}
    \begin{subfigure}[b]{0.47\textwidth}
         \centering
         \includegraphics[width=\textwidth]{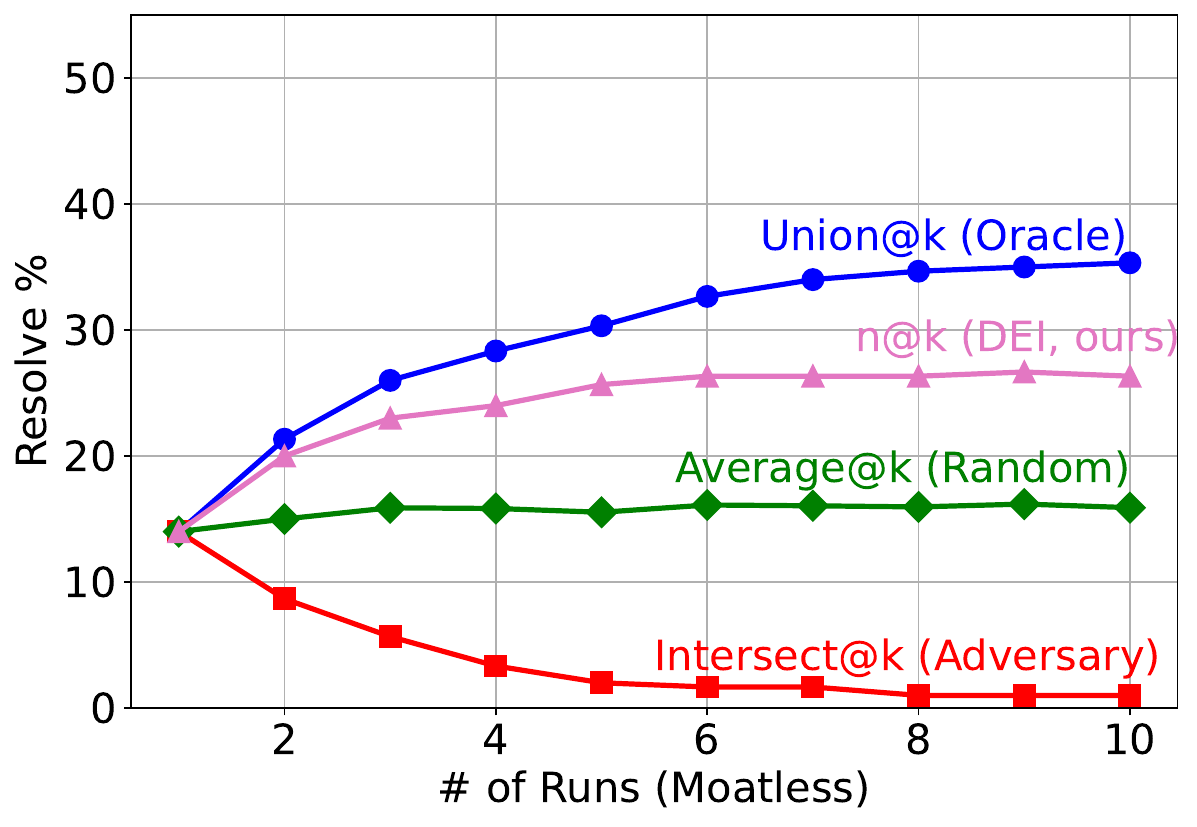}
         \label{fig:three sin x}
     \end{subfigure}

    \vspace{-5pt}
    \caption{How different metrics change as more candidate solutions are involved. In all 4 scenarios, there is a huge gap between Union@k and Average@k.}%
    \label{fig:diversity_analysis}
\end{figure}

To answer this question, we report the ``@k'' metrics of 10 different agents and 10 runs of single agents in \autoref{fig:diversity_analysis}.
The detailed values of these metrics can also be found in \autoref{table:multirun}.

Several observations can be made about the results:

\textbf{SWE agents resolve very different sets of issues across agents and agent runs. Their full potential is far from fully released.}
In all four subfigures, the gap between Union@k and Average@k, as well as between Average@k and Intersect@k, is large.
As $k$ -- the number of candidates -- gets larger, the gap also gets larger.
For 2 of the 4 settings, Union@k is more than 2x larger than Average@k for $k=10$.
The other 2, Union@k is more than 1.5x larger than Average@k for $k=10$.
This indicates that current SWE agents are potentially capable of resolving a lot more issues, as long as we have a reranker that can tell which candidates are correct.

\textbf{Different agents resolve more distinct issues than different runs of a single agent. In other words, diversity does empower intelligence.}
The absolute/relative difference between Union@k and Average@k is much larger in the first subfigure than in the following three subfigures.
For the ``10 different agents'' setting, as $k$ approaches 10, the distinct issues resolved are $2\times $ the average number of issues resolved by a single agent in the group.

\begin{table}[htb]
\centering
\scriptsize
\resizebox{\textwidth}{!}{%
\begin{threeparttable}
\caption{\small Resolve rates of top-ranking submissions on SWE-Bench Lite. We evaluate 3 DEI Committees formed by different groups of agents. Each DEI Committee outperforms the best agent in it significantly. \baseline-Open, a committee formed by 4 open-source agents can beat many closed-source agents.}
\label{table:comparison}
\begin{tabular}{ccccccc}

\toprule
\textsc{\textbf{Dei}} Group & \% Resolve  & System             & Open Src & Trajs & Open 
Candidates & Backend LLM                \\
\midrule
\rowcolor{heavyblue}\textbf{1} & \textbf{55.0}           & Salesforce Research \baseline-1              & \cmark         & \cmark        & \xmark                     & gpt4o                     \\
\rowcolor{lightblue} 1 & 50.6         & Cosine Genie       & \xmark         & \xmark          & -                      & ``Fine-tuned OpenAI''          \\
\rowcolor{lightblue} 1 & 43.0         & CodeStory Aide     & \xmark          & \xmark                & -                      & gpt4o, Claude 3.5 Sonnet \\
- & 38.0         & AbenteAI MentatBot & \xmark          & \xmark               & -                     & gpt4o          \\
\rowcolor{heavyorange}\textbf{2} & \textbf{37} & Salesforce Research \baseline-2 & \cmark & \cmark & \xmark & gpt4o\\
\rowcolor{heavygreen}\textbf{Open} & \textbf{34.3} & Salesforce Research \baseline-Open & \cmark & \cmark & \cmark & gpt4o \\
- & 34.0 & Bytedance MarsCode & \xmark & \xmark & - & gpt4o \\
- & 33.0 & Alibaba Lingma & \xmark\tnote{1} & \xmark & - & gpt-4-1106-preview \\
\rowcolor{lightorange}2 & 31.3 & Factory Code Droid & \xmark & \xmark & - & ``Anthropic and OpenAI'' \\
\rowcolor{lightorange}2 & 30.6 & AutoCodeRover & \xmark\tnote{2} & \xmark & - & gpt4o \\ 
\rowcolor{lightorange}2 & 29.6 & Amazon Q Dev. & \xmark & \xmark & - & Unknown \\
\rowcolor{lightorange}2 & 28.3 & CodeR & \xmark\tnote{1} & \xmark & - & gpt-4-1106-preview \\
\rowcolor{lightorange}2 & 28.0 & MASAI & \xmark\tnote{1} & \xmark & - & Unknown \\
- & 27.6 & SIMA & \xmark\tnote{1} & \cmark & \cmark\tnote{3} & gpt4o \\
\rowcolor{lightgreen}Open & 27.3 & Agentless & \cmark & \cmark & - & gpt4o \\
\rowcolor{lightgreen}Open & 26.6 & Moatless Tools & \cmark & \cmark & - & Claude 3.5 Sonnet \\
- & 26.6 & IBM Research Agent & \xmark & \xmark & - & Unknown \\
\rowcolor{lightgreen}Open & 26.3 & Aider & \cmark & \xmark & - & gpt4o, Claude 3 Opus \\
\rowcolor{lightgreen}Open & 26.0 & OpenDevin + CodeAct & \cmark & \cmark & - & gpt4o \\
\bottomrule 

\end{tabular}
\begin{tablenotes}
\item[1]Their repo has no code yet.
\item[2]An earlier version is open-source. The current one is not.
\item[3]Candidates are generated by a ``modification of moatless tools''.
\end{tablenotes}

\end{threeparttable}
}
\end{table}

\subsubsection{Research Question 2: How much does \method help?}

We apply \baseline to the candidates in \autoref{fig:diversity_analysis} as they are added to the group.
Our findings are:

\textbf{\baseline helps in most cases.} For most values of $k$ in all subfigures, we observe a significant improvement of n@k over Average@k, indicating that \baseline selects correct candidates much better than a random baseline.

\textbf{\baseline helps more when the candidates come from different agents.}
This finding resonates with a similar finding from research question one:
Since candidates from multiple agents have a larger potential for improvement (Union@k - Average@k), the actual improvements created by \baseline (n@k - Average@k) are also larger.
This suggests that given a limited budget of candidates, it would be better to choose a diversity of agents over multiple runs of the same agent.

\textbf{As $k$ gets larger, \baseline's improvement first increases and then plateaus.} While larger $k$ generally indicates higher n@k, the margin gets smaller and there are cases when an increase in $k$ results in a slight drop in performance. This suggests that the current \baseline is not ideal for a large group of agents and there is still room for a better reranking mechanism.

Based on the lessons above, we propose three \baseline groups in which each candidate is from a different agent and no more than 5 candidates exist for each instance.
The members of these \baseline groups and their performance are reported in \autoref{table:comparison}.
\baseline-1 consists of the top 2 agents.
\baseline-2 consists of 5 closed-source agents that have high performance on the leaderboard.
\baseline-Open consists of 4 open-source agents so that we know future researchers can run the entire pipeline.
As \autoref{table:comparison} shows,
all three \baseline instances outperform the best candidate in the group.
Surprisingly, \baseline-Open shows a 7\% increase in resolve rates and beats most of the closed-source systems.

\begin{table}[ht]
\centering
\scriptsize
\begin{threeparttable}
\caption{\small How different metrics change as more candidate solutions are involved. As the number of candidates $k$ gets larger, the improvement from DEI also increases significantly.}
\label{table:multirun}
\begin{tabular}{llccccc}
\toprule
System & $k$ & Intersect@k & Average@k & 1@k (DEI, ours) & Improvement from DEI & Union@k \\
\midrule
\multirow{10}{*}{\textbf{10 Agents}} &1&26.7&26.7&26.7&+0.0&26.7\\
&2&18.3&27.3&28.7&+1.3&36.3\\
&3&15.7&27.7&32.3&+4.7&42.3\\
&4&13.0&27.3&34.0&+6.7&45.0\\
&5&10.3&27.3&35.0&+7.7&48.3\\
&6&8.3&26.9&34.3&+7.4&49.0\\
&7&6.7&26.8&35.3&+8.6&52.0\\
&8&6.7&26.9&34.7&+7.8&52.7\\
&9&6.3&27.3&35.7&+8.4&53.3\\
&10&4.7&26.6&35.7&+9.1&54.3\\

\midrule
\multirow{10}{*}{\textbf{10 Runs from Agentless}}&1&20.3&20.3&20.3&+0.0&20.3\\
&2&14.3&19.5&22.0&+2.5&24.7\\
&3&13.7&20.3&24.0&+3.7&28.0\\
&4&12.0&20.3&22.0&+1.7&29.3\\
&5&11.7&20.5&23.7&+3.2&31.3\\
&6&11.0&20.4&23.7&+3.3&32.3\\
&7&10.7&20.2&24.3&+4.1&33.0\\
&8&9.3&20.0&25.7&+5.6&34.0\\
&9&9.0&20.4&26.0&+5.6&34.0\\
&10&9.0&20.4&26.0&+5.6&34.7\\
\midrule
\multirow{10}{*}{\textbf{10 Runs from Aider}}&1&22.3&22.3&22.3&+0.0&22.3\\
&2&17.7&22.3&22.3&+0.0&27.0\\
&3&15.7&21.6&22.0&+0.4&29.0\\
&4&14.7&22.2&25.0&+2.8&32.3\\
&5&14.0&22.1&25.3&+3.2&35.3\\
&6&13.0&21.8&25.3&+3.6&35.3\\
&7&12.0&21.7&24.3&+2.6&36.0\\
&8&12.0&21.8&24.3&+2.5&37.0\\
&9&11.7&21.7&24.3&+2.6&37.7\\
&10&10.7&21.7&24.7&+3.0&38.0\\
\midrule
\multirow{10}{*}{\textbf{10 Runs from Moatless}}&1&14.0&14.0&14.0&+0.0&14.0\\
&2&8.7&15.0&20.0&+5.0&21.3\\
&3&5.7&15.9&23.0&+7.1&26.0\\
&4&3.3&15.8&24.0&+8.2&28.3\\
&5&2.0&15.5&25.7&+10.1&30.3\\
&6&1.7&16.1&26.3&+10.2&32.7\\
&7&1.7&16.0&26.3&+10.3&34.0\\
&8&1.0&16.0&26.3&+10.4&34.7\\
&9&1.0&16.2&26.7&+10.5&35.0\\
&10&1.0&15.9&26.3&+10.4&35.3\\

\bottomrule
\end{tabular}

\end{threeparttable}
\end{table}

\subsection{Ablation and Analyses}

In this subsection, we demonstrate some ablation studies to investigate the effectiveness of different components in the framework, in order to answer the following questions.
To advocate for open science, all the ablation experiments are conducted on either our own reproduction of open-source SWE agents or their official generations.

\textbf{Question 1: Does \method get better with more votes?}

\textbf{Answer 1: Yes.} Arguably, \method itself has the same potential characteristics as SWE agents that may cause diverse outputs.
So it is important for us to harness the diverse outputs of \method as well.
However, unlike SWE agents whose outputs are code patches, \method's output is an integer score, which can easily be aggregated and averaged.
This is why we give \method more votes and rerank the candidates according to the average of scores.
In most \baseline experiments, we allow 10 votes for each candidate patch.
To investigate whether more votes lead to better patch reviewing, we directly take the scores generated for \baseline-Open, \baseline-Agentless, \baseline-Aider, and \baseline-Moatless, and check for various values of $m$, how the first $m$ scores can help us find the best patch.

\begin{figure}[t]
     \centering
     \begin{subfigure}[b]{0.47\textwidth}
         \centering
         \includegraphics[width=\textwidth]{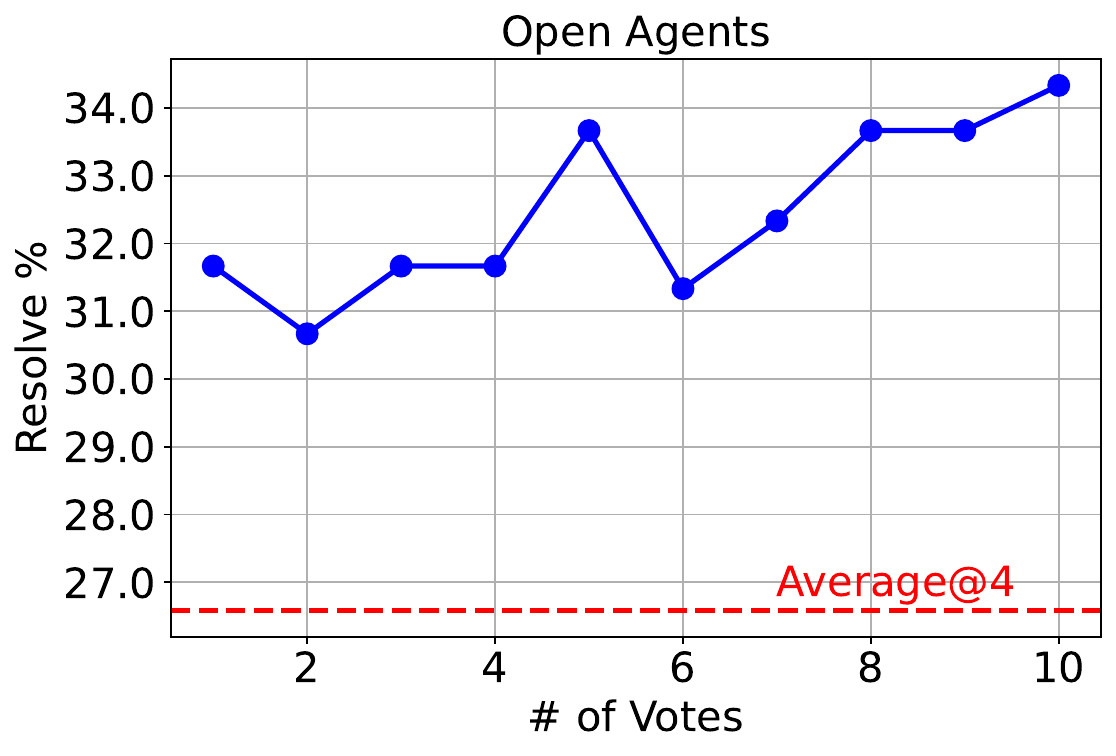}
         \label{fig:multiagent}
     \end{subfigure}
     \begin{subfigure}[b]{0.47\textwidth}
         \centering
         \includegraphics[width=\textwidth]{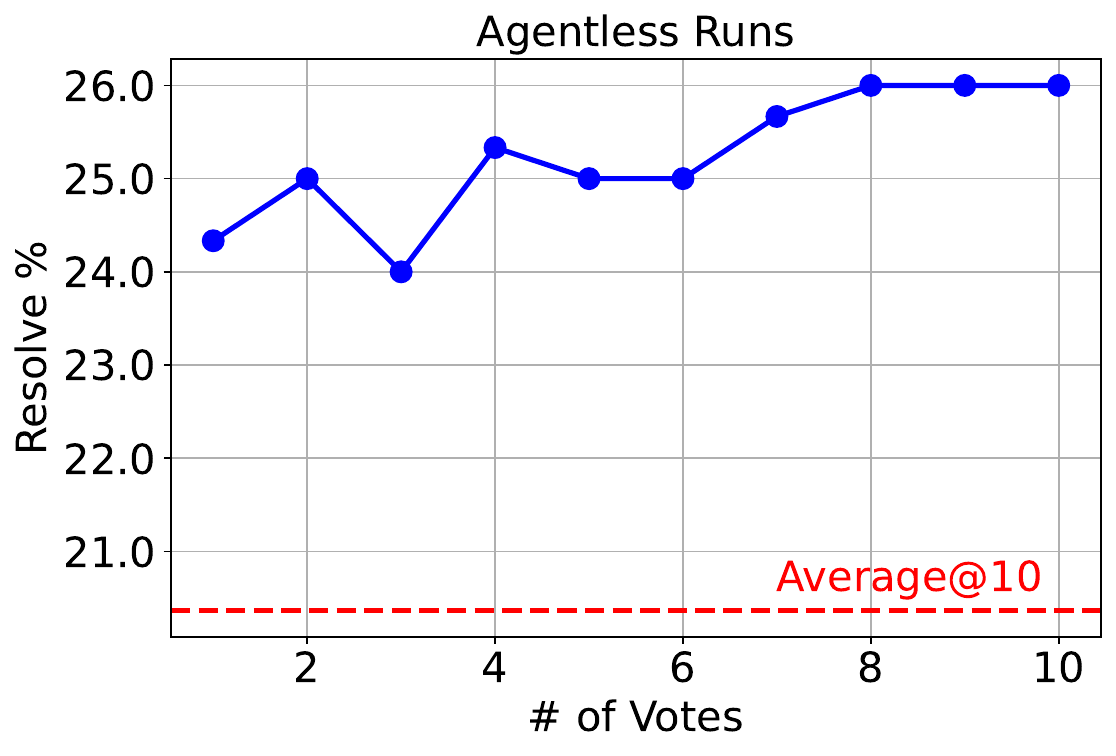}
         \label{fig:gaga}
     \end{subfigure}
    \vspace{-10pt}
    \\
    \begin{subfigure}[b]{0.47\textwidth}
         \centering
         \includegraphics[width=\textwidth]{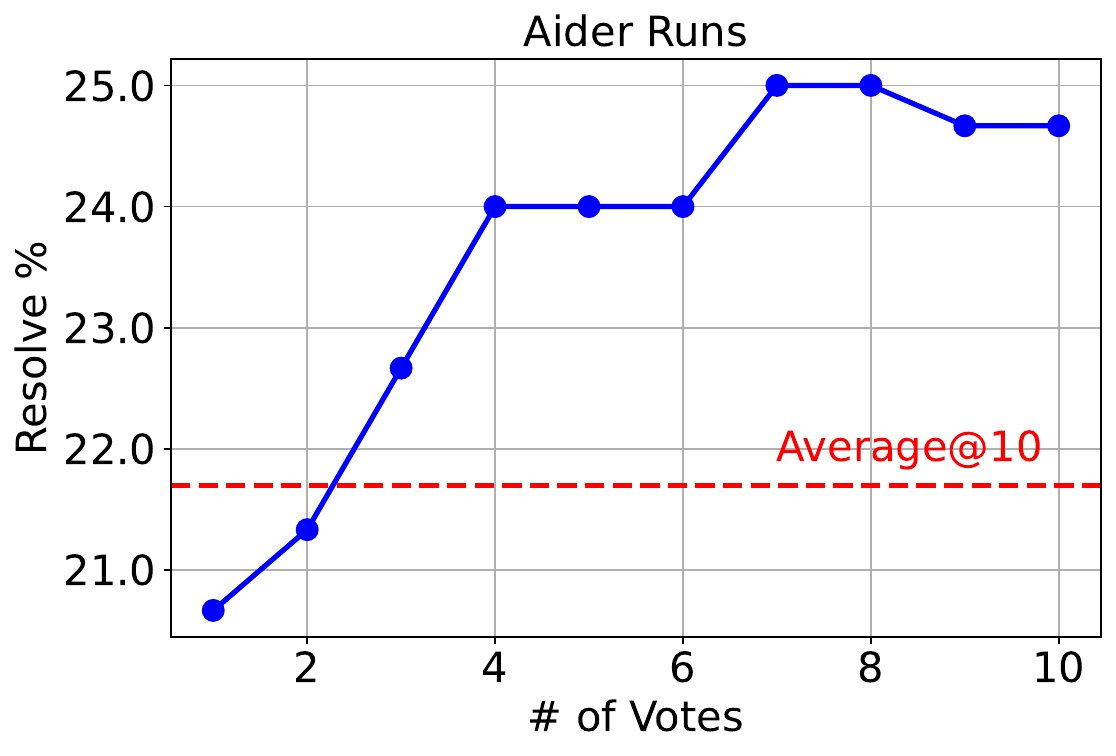}
         \label{fig:three sin x}
     \end{subfigure}
    \begin{subfigure}[b]{0.47\textwidth}
         \centering
         \includegraphics[width=\textwidth]{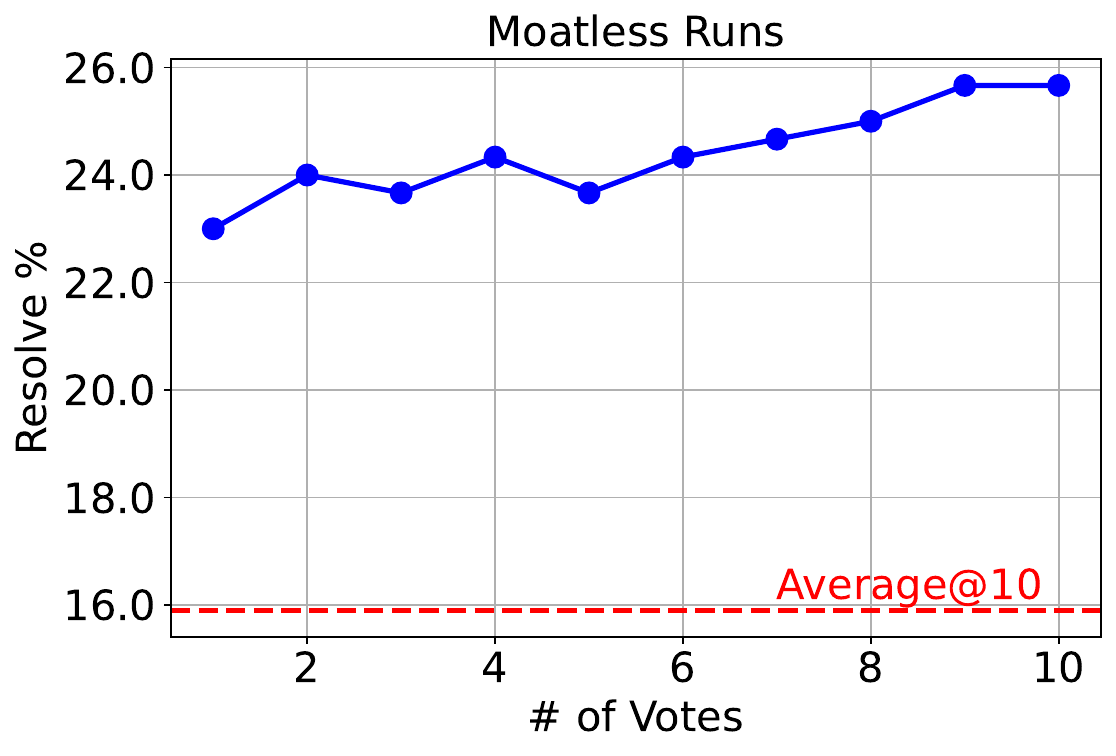}
         \label{fig:three sin x}
     \end{subfigure}

    \vspace{-5pt}
    \caption{How the performance of \baseline changes as the LLM is given more votes for scoring.}%
    \label{fig:votes_analysis}
\end{figure}

As demonstrated in \autoref{fig:votes_analysis}, more votes generally lead to better resolve rates.
Another finding is that for 3 out of the 4 evaluation settings, \baseline was able to get much better performance than the average candidate with only one vote.
Even when \baseline wasn't able to get better than average with one vote, it managed to get an improvement with only three votes.
These results suggest that \baseline itself also produces diverse outputs, but it is easier to aggregate them via score averaging.

\textbf{Question 2: Are the explanations necessary?}

\textbf{Answer 2: Yes.} We remove the part about asking for explanations from the prompt and compare \baseline-Open, \baseline-Agentless, \baseline-Aider, and \baseline-Moatless under the same evaluation setting with and without explanations. We report their resolve rates in \autoref{tab:expl}. For all 4 settings we evaluated, \baseline with explanations performs slightly better than \baseline without explanations.

\begin{table}[ht]
\caption{Comparing \baseline's resolve rates with and without explanations.}
\centering
\small
\begin{tabular}{lcccc}
\toprule
                   & Open Agents & Agentless & Aider & Moatless \\
\midrule
\baseline w/ expl.   & 34.6  &  26.0  &  24.6     &   25.6       \\
\baseline w/ o expl. & 32.3  &  23.0   & 23.3      & 25.3 \\
\bottomrule
\end{tabular}
\label{tab:expl}
\end{table}

\section{Conclusion}
In this paper, we present Diversity Empowered Intelligence (\method), a meta-policy module designed to integrate with any existing SWE agent frameworks to enable scalable management and collaboration among specialized agents, thereby fostering a more powerful software engineering organization. Through extensive evaluations, we find that different agents show a great level of diversity in the issues they resolve: a group of agents with an average resolve rate of 26.6\% can actually solve 54.3\% of the issues if we have an oracle that selects the correct candidate.
\method, as our first step towards harnessing such diversity, can improve the group's resolve rate to 34.3\% (+7\%), suggesting that LLMs are great code reviewers.
These findings mirror the benefits of diversity in the tech industry, where diverse perspectives and skills lead to greater innovation and problem-solving capabilities.

\paragraph{Broader Impacts.}
\method represents our initial step toward realizing a fully automated organizational AI. We believe that the full potential of multi-agent AI systems extends beyond enhancing task completion accuracy with agentic workflows, which is the current focus of most industry practices. Instead, \method offers a horizontal, scaling-out approach that facilitates the collaboration and integration of existing diverse agents without necessitating refactoring of engineering work. This capability not only optimizes and speeds up immediate software development tasks but also sets the groundwork for future innovations in AI-driven organizational management.

\bibliography{iclr2024_conference}
\bibliographystyle{iclr2024_conference}

\clearpage
\appendix
\section{Appendix}

\subsection{Agents Evaluated} \label{app:agents}

We add the following agents to the \method Committee (the one in Figure 3) in the order they appear (the order is generated by randomly shuffling their chronological order using python's random shuffle function with a random seed of 42):

\begin{enumerate}
    \item \texttt{20240612 IBM Research Agent101}
    \item \texttt{20240612 MASAI gpt4o}
    \item \texttt{20240604 CodeR}
    \item \texttt{20240523 aider}
    \item \texttt{20240630 agentless gpt4o}
    \item \texttt{20240617 moatless gpt4o}
    \item \texttt{20240725 opendevin codeact v1.8 claude35sonnet}
    \item \texttt{20240706 sima gpt4o}
    \item \texttt{20240621 autocoderover-v20240620}
    \item \texttt{20240509 amazon-q-developer-agent-20240430-dev}
\end{enumerate}

\end{document}